\documentclass[a4paper]{jpconf}
\usepackage{graphicx}
\usepackage{amsmath}
\usepackage{amssymb}

\renewcommand{\arraystretch}{1.6}
\newcommand{\hc}{\mathrm{h.c.}}

\newcommand{\id}{\mathbb{I}}

\begin{document}

\title{Master Majorana neutrino mass parametrization}

\author{Isabel Cordero-Carri\'on$^1$, Martin Hirsch$^2$ and Avelino Vicente$^2$}

\address{$^1$ Departamento de Matem\'aticas, Universitat de Val\`encia, E-46100 Burjassot, 
Val\`encia, Spain}
\address{$^2$ Instituto de F\'{\i}sica Corpuscular (CSIC-Universitat de Val\`{e}ncia),
Apdo. 22085, E-46071 Valencia, Spain}

\ead{avelino.vicente@ific.uv.es}

\begin{abstract}
After showing that the neutrino mass matrix in all Majorana models can
be described by a general master formula, we will present a master
parametrization for the Yukawa matrices, also valid for all Majorana
models, that automatically ensures agreement with neutrino oscillation
data. The application of the master parametrization will be
illustrated in an example model.
\end{abstract}

\section{Introduction}
\label{sec:intro}

The existence of non-zero neutrino masses is nowadays an established
experimental fact that calls for an extension of the Standard Model
(SM) of particle physics. In fact, many neutrino mass models have been
proposed, see
\cite{Ma:2016mwh,CentellesChulia:2018gwr,Ma:1998dn,Cai:2017jrq,Cepedello:2018rfh,Boucenna:2014zba,Anamiati:2018cuq}
for some recent reviews and classification papers.

Here we will concentrate on Majorana neutrino mass models. We will
first show that in this class of models the neutrino mass matrix can
always be regarded as a particular case of a master formula. This
general expression is written in terms of generic mass and Yukawa
matrices which take specific forms in a given model. We will then
enforce the agreement with neutrino oscillation data by introducing a
master parametrization of the Yukawa matrices appearing in this
formula. In order to illustrate the application of this
parametrization we will consider an example in the BNT model
\cite{Babu:2009aq}, a model that requires one to use the full power of
the master parametrization. For more details on the master formula and
parametrization, we refer to \cite{Cordero-Carrion:2018xre} as well as
to the extended work \cite{future}.

\section{The master formula}
\label{sec:master}

A Majorana neutrino mass matrix can always be written as
\begin{equation} \label{eq:master}
m = f \, \left( y_1^T \, M \, y_2 + y_2^T \, M^T \, y_1 \right) \, .
\end{equation}
Here $m$ is the neutrino mass matrix, a $3 \times 3$ complex symmetric
matrix that can be diagonalized as
\begin{equation} \label{eq:m-U}
D_m = \text{diag} \left( m_1, m_2, m_3 \right) = U^T \, m \, U \, ,
\end{equation}
with $U$ a $3 \times 3$ unitary matrix. $y_1$ and $y_2$ are two
general $n_1 \times 3$ and $n_2 \times 3$ complex Yukawa matrices,
respectively, and $M$ is a $n_1 \times n_2$ complex matrix with
dimension of mass. In the following we will assume $n_1 \geq
n_2$. Since $m$ must contain at least two non-vanishing eigenvalues in
order to accommodate the solar and atmospheric mass scales, $r_m =
\text{rank}(m) = 2 \, \text{or} \, 3$.

Eq. \eqref{eq:master} is a \textit{master formula} valid for all
Majorana neutrino mass models. In fact, the resulting neutrino mass
matrices in specific models can be seen as particular cases of this
general expression. Let us consider three examples:

\begin{itemize}
\item In the {\bf type-I seesaw} with $3$ generations of right-handed
  neutrinos, the light neutrino mass matrix is given by the well-known
  seesaw formula, $m = - \langle H^0 \rangle^2 \, y^T M_R^{-1}
  y$. This can be obtained with the master formula by taking $n_1 =
  n_2 = 3$ and the specific values $f=-1$, $y_1 = y_2 = y/\sqrt{2}$
  and $M = \langle H^0 \rangle^2 \, M_R^{-1}$, with $\langle H^0
  \rangle = v / \sqrt{2}$ the SM Higgs ($H$) vacuum expectation value (VEV)
  and $M_R$ the Majorana mass matrix for the right-handed neutrinos.
\item The {\bf inverse seesaw} \cite{Mohapatra:1986bd} would correspond to
  the same $y_{1,2} = y$ and $f$ values, but $M = \langle H^0
  \rangle^2 \, (M_R^T)^{-1} \mu M_R^{-1}$, with $\mu$ the small lepton
  number violating parameter.
\item In the {\bf scotogenic model} \cite{Ma:2006km}, the neutrino
  mass matrix is induced at the 1-loop level and can also be seen as a
  particular case of the general master formula. It corresponds to $f
  = \lambda_5 / (16 \pi^2)$ and $M = \langle H^0 \rangle^2 \, M_R^{-1}
  F_{\rm loop}$, with $\lambda_5 \, (H^\dagger \eta)^2$ the quartic
  term involving the usual and inert ($\eta$) scalar doublets, and
  $F_{\rm loop}$ a matrix containing loop functions.
\end{itemize}

Finally, a non-trivial example with with $y_1 \ne y_2$ will be
considered in Sec. \ref{sec:example}.

\section{The master parametrization}
\label{sec:param}

In order to guarantee consistency with neutrino oscillation data, the
Yukawa matrices $y_1$ and $y_2$ in Eq. \eqref{eq:master} can be written as
\begin{align}
y_1 & = \frac{1}{\sqrt{2 \, f}} \, V_1^\dagger \, \left( \begin{array}{c}
\Sigma^{-1/2} \, W \, A \\ X_1 \\ X_2 \end{array} \right) \, 
\bar{D}_{\sqrt{m}} \, U^\dagger \, , \label{eq:par1} \\
y_2 & = \frac{1}{\sqrt{2 \, f}} \, V_2^\dagger \, \left( \begin{array}{c}
\Sigma^{-1/2} \, \widehat W^\ast \, \widehat B \\ X_3 \end{array} \right) 
\, \bar{D}_{\sqrt{m}} \, U^\dagger \, . \label{eq:par2}
\end{align}
This is the \emph{master parametrization}. We now proceed to define
the matrices that appear in Eqs. \eqref{eq:par1} and
\eqref{eq:par2}. First, we have introduced the diagonal matrix
$\bar{D}_{\sqrt{m}}$, given by
diag$\left(\sqrt{m_1},\sqrt{m_2},\sqrt{m_3}\right)$ if $r_m=3$ or
diag$\left(\sqrt{m_1},\sqrt{m_2},\sqrt{v}\right)$ if $r_m=2$. The
matrix $M$ has been singular-value decomposed as
\begin{equation} \label{eq:M-SVD}
M = V_1^T \, \widehat \Sigma \, V_2 \, ,
\end{equation}
where $\widehat \Sigma$ is a $n_1 \times n_2$ matrix that can be written as
\begin{equation}
\widehat \Sigma = \left(\begin{array}{c} \begin{array}{cc} 
\Sigma & 0 \\ 0 & 0_{n_2-n} \end{array} \\ \hline 0_{n_1-n_2} 
\end{array}\right) \, ,
\end{equation}
and $\Sigma =
\text{diag}\left(\sigma_1,\sigma_2,\dots,\sigma_n\right)$ is the
diagonal $n \times n$ matrix that contains the positive and real
singular values of $M$ ($\sigma_i>0$).  $V_1$ and $V_2$ are two
unitary matrices, with dimensions $n_1 \times n_1$ and $n_2 \times
n_2$, respectively. $X_1$, $X_2$ and $X_3$ are three arbitrary complex
matrices with dimensions $(n_2-n)\times 3$, $(n_1-n_2)\times 3$ and
$(n_2-n)\times 3$, respectively, and whose entries have dimensions of
mass$^{-1/2}$. $\widehat W$ is an $n\times n$ matrix defined as
\begin{equation}
\widehat W = \left(W \quad \bar{W}\right) \, ,
\end{equation}
where $W$ is an $n\times r$ complex matrix, such that $W^\dagger W =
W^T W^* = \id_r$. Here we have defined $r=\text{rank}(W)$. The matrix
$\bar{W}$ is an $n\times(n-r)$ complex matrix, built with vectors that
complete those in $W$ to form an orthonormal basis of
$\mathbb{C}^n$. Furthermore, $A$ is an $r\times 3$ matrix that can be
expressed as
\begin{equation}
A = T \, C_1 \, ,
\end{equation}
with $T$ an upper-triangular $r\times r$ invertible square matrix with
$T_{ii} > 0$, and $C_1$ is an $r\times 3$ matrix. Finally, $\widehat
B$ is an $n\times 3$ complex matrix, which can be written in blocks as
\begin{equation}
\widehat B = \left( \begin{array}{c} B \\ \bar{B} \\ \end{array} \right) \, ,
\end{equation}
with $\bar{B}$ an arbitrary $(n-r)\times 3$ complex matrix and $B$ an
$r\times 3$ complex matrix given by
\begin{equation} \label{eq:Bexp}
B \equiv B\left( T , K , C_1 , C_2 \right) 
= \left( T^T \right)^{-1} \, \left[ C_1 \, C_2 + K \, C_1 \right] \, .
\end{equation}
In the last equation we have introduced the antisymmetric $r\times r$
square matrix $K$ and the $3\times 3$ matrix $C_2$. The form of the
matrices $C_1$ and $C_2$ depends on the ranks $r_m$ and $r$ (see
\cite{future} for all the expressions). For instance, for $r_m = r =
3$ these matrices are given by
\begin{align}
C_1 = \id_3, \quad
C_2 = \id_3 + K_{12} \, \frac{T_{13}}{T_{11}} \, \left( \begin{array}{ccc}
0 & 0 & 0 \\
0 & 0 & 1 \\
0 & -1 & 0 \end{array} \right) \, . \label{eq:C1C2}
\end{align}

The use of the master parametrization might look complicated but is
actually straightforward. The first step is to use information from
neutrino oscillation experiments (typically from a global fit) to fix
the light neutrino masses and leptonic mixing angles appearing in
$\bar D_{\sqrt{m}}$ and $U$, respectively. Next, one must compare the
expression for the neutrino mass matrix in the specific model under
study with the general master formula in Eq. \eqref{eq:master}. This
way one identifies the global factor $f$, the Yukawa matrices $y_1$
and $y_2$ and the matrix $M$, and by singular-value decomposing the
latter one determines $\Sigma$, $V_1$ and $V_2$. Finally, one can
randomly scan over the free parameters contained in the matrices
$\widehat W$, $X_{1,2,3}$, $\bar B$, $T$, $K$ and $C_{1,2}$ to compute
the Yukawa matrices $y_1$ and $y_2$ by means of Eqs. \eqref{eq:par1}
and \eqref{eq:par2}.

Let us now compare to the Casas-Ibarra
parametrization~\cite{Casas:2001sr}. As explained above, the master
parametrization can be applied to any Majorana neutrino mass model,
while the use of the Casas-Ibarra parametrization is restricted to the
type-I seesaw (and similar models). Therefore, they should agree in
that case. First, we remind the reader that comparing the neutrino
mass matrix in this model to our master formula one finds $y_1 = y_2 =
y/\sqrt{2}$, $n_1 = n_2 = n = r = 3$, $f=-1$ and $M = \langle H^0
\rangle^2 M_R^{-1}$. Since $M$ is symmetric, it can be diagonalized by
a single matrix, and hence $V_1 = V_2$. Moreover, this matrix can be
taken to be the identity when the right-handed neutrinos are given in
their mass basis. Finally, since $n_1 = n_2 = n = r = 3$ the matrices
$X_{1,2,3}$, $\overline W$ and $\overline B$ just drop from all the
expressions. The condition $y_1 = y_2$ can be shown to be equivalent
to $W^T W A = B$, which in turn leads to $B = \left(A^T\right)^{-1}$
and $R = W \, A$, with $R$ a $3 \times 3$ orthogonal matrix. With
these ingredients at hand one can simply use Eqs. \eqref{eq:par1} and
\eqref{eq:par2} to find
\begin{equation}
y = \sqrt{2} \, y_1 = \sqrt{2} \, y_2 = i \, \Sigma^{-1/2} \, R \, D_{\sqrt{m}} \, U^\dagger \, ,
\end{equation}
which, after identifying $R$ with the usual Casas-Ibarra matrix, is
nothing but the Casas-Ibarra parametrization
\cite{Casas:2001sr}. Therefore, we see that the Casas-Ibarra
parametrization can be interpreted as a particular case of the master
parametrization.

\section{An example application}
\label{sec:example}

Finally, we would like to show an application of the master
parametrization to the BNT model \cite{Babu:2009aq}. The particle
content of this model includes three generations of the vector-like
fermions $\psi_{L,R}$, which transform as $({\bf 1}, {\bf 3}, -1)$
under the SM gauge group and the scalar $\Phi$, which transfors as
$({\bf 1}, {\bf 4}, 3/2)$. The quantum numbers of the new particles in
the BNT model are given in Table~\ref{tab:BNT}.

{
\renewcommand{\arraystretch}{1.4}
\begin{table}
\centering
{\setlength{\tabcolsep}{0.5em}
\begin{tabular}{| c c c c c |}
\hline  
 & generations & $\mathrm{SU(3)}_c$ & $\mathrm{SU(2)}_L$ & $\mathrm{U(1)}_Y$ \\
\hline
\hline    
$\Phi$ & 1 & ${\bf 1}$ & ${\bf 4}$ & $3/2$ \\
\hline
\hline    
$\psi_{L,R}$ & 3 & ${\bf 1}$ & ${\bf 3}$ & $-1$ \\ 
\hline
\end{tabular}
}
\caption{New particles in the BNT model.}
\label{tab:BNT}
\end{table}
}

\begin{figure}[t]
\centering
\includegraphics[width=0.6\textwidth]{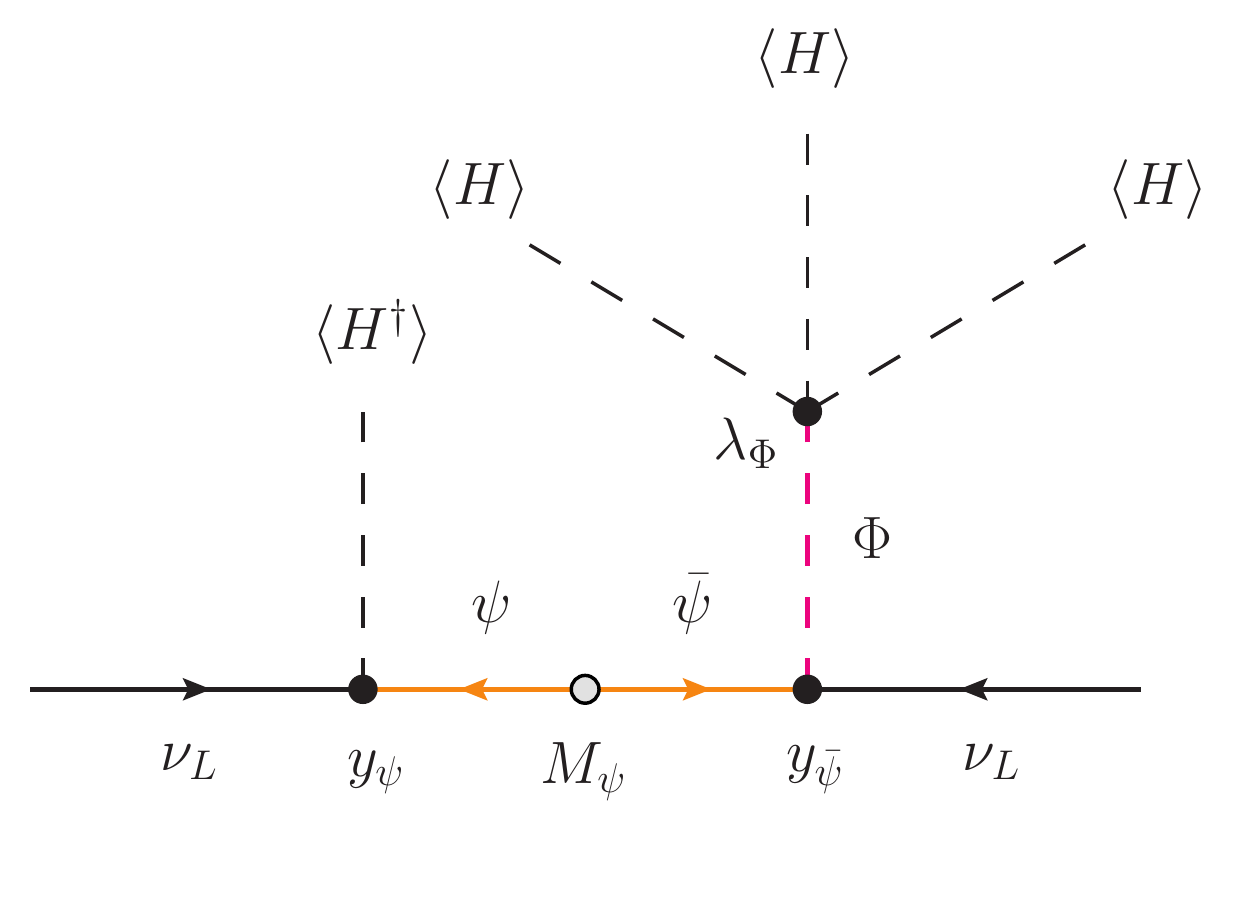}
\caption{Neutrino mass generation in the BNT model.} 
\label{fig:BNTmodel}
\end{figure}

The Lagrangian contains the following terms
\begin{align} \label{eq:LagBNT}
-\mathcal L &\supset y_\psi \, \overline{L} \, H \, \psi_R + y_{\bar \psi} \, \overline{L^c} \, \Phi \, \psi_L + M_\psi \overline \psi \, \psi + \lambda_\Phi \, H^3 \, \Phi^{\dagger} + \hc \, ,
\end{align}
where we have omitted gauge and flavor indices for the sake of
clarity. In the presence of a non-zero $\lambda_\Phi$ coupling the
model breaks lepton number in two units and induces neutrino masses as
shown in Fig.~\ref{fig:BNTmodel}. The resulting neutrino mass matrix
is given by
\begin{align} \label{eq:mnuBNT}
m = \frac{\lambda_\Phi v^4}{4 M_\Phi^2} \, \left[ y_\psi^T \, M_\psi^{-1} \, y_{\bar \psi} + y_{\bar \psi}^T \, (M_\psi^{-1})^T \, y_{\psi} \right] \,.
\end{align}

Furthermore, the $\lambda_\Phi$ term induces a non-zero VEV for the
neutral component of $\Phi$, $\Phi^0$,
\begin{equation}
\langle \Phi^0 \rangle = \frac{v_{\Phi}}{\sqrt{2}} = \frac{\lambda_\Phi v^3}{2 \sqrt{2} M_\Phi^2} \, .
\end{equation}

One cannot apply the Casas-Ibarra parametrization in the BNT model
since one has two independent $y_1 = y_\psi$ and $y_2 = y_{\bar \psi}$
Yukawa matrices. Therefore, the master parametrization is required
in order to guarantee consistency with neutrino oscillation experiments. First, we compare to 
Eq. \eqref{eq:master} and identify
\begin{equation}
f = \frac{\lambda_\Phi v^2}{2 M_\Phi^2} \, , \quad M = \frac{v^2}{2} \, M_\psi^{-1} \, .
\end{equation}
Moreover, the matrices $y_1$, $y_2$ and $M$ are $3 \times 3$ in this
model and then $n_1 = n_2 = 3$. One also has $n = 3$ and $\widehat
\Sigma \equiv \Sigma$. Finally, we consider the choice $r = r_m = 3$,
implying that the matrices $X_{1,2,3}$ and $\bar B$ are absent, while
$C_1$ and $C_2$ are given in Eq. \eqref{eq:C1C2}.

\begin{figure}[t]
 \centering
 \includegraphics[width=0.6\textwidth]{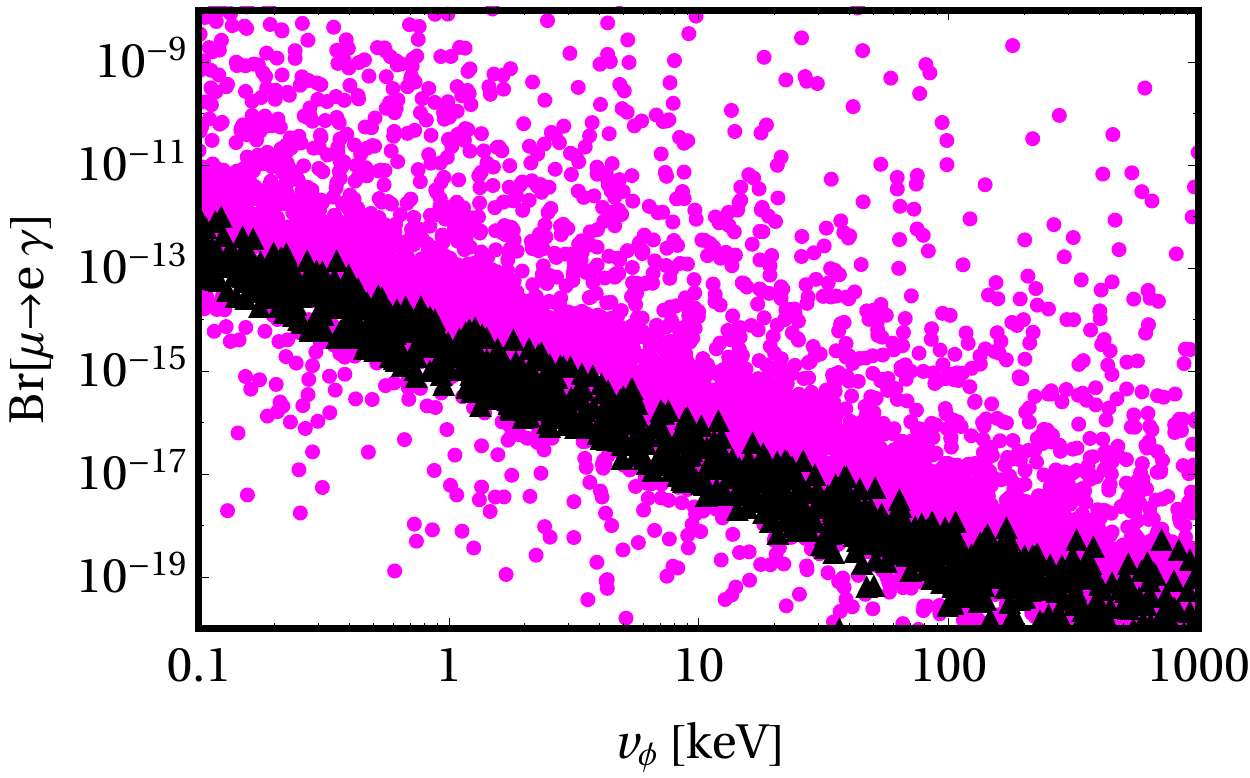}
 \caption{Br$(\mu\to e\gamma)$ as a function of $v_{\Phi}$ in the BNT
   model. Neutrino oscillation parameters are allowed to vary within
   the 3 $\sigma$ ranges determined in \cite{deSalas:2017kay},
   assuming normal hierarchy. $M_{\Psi}$ has been randomly taken in
   the interval $[0.5,2]$ TeV and $W$ fixed to the identity
   matrix. The purple points correspond to a scan in which the
   elements of the matrices $T$ and $K$ are randomly taken in the
   following ranges: $T_{ii} \in [0,2]$ and $K_{ij}$, $T_{ij}$ (with
   $i\ne j$) $\in [-1,1]$. The black points correspond to a simplified
   scan with $T = \id$ and $K = 0$.
\label{fig:Ex1}}
\end{figure}

We have performed numerical scans to show the usefulness of the master
parametrization. In order to do that we have made use of the neutrino
oscillation parameters derived by the global fit
\cite{deSalas:2017kay}, implemented the model in {\tt SARAH}
\cite{Staub:2013tta} and obtained numerical results with {\tt SPheno}
\cite{Porod:2011nf}. We show a selected result on the lepton flavor
violating observable Br$(\mu\to e\gamma)$, computed with the {\tt
  FlavorKit} package \cite{Porod:2014xia}, in Fig. \ref{fig:Ex1}. This
figure serves to illustrate a crucial point when running a numerical
scan. One can take simple forms for the matrices that appear in the
master parametrization (for instance, $T = \id$ or $K = 0$). However,
that would cover a limited region of the parameter space of the model,
potentially leading to fictitious correlations that get broken in
other parameter regions. Fig. \ref{fig:Ex1} precisely shows the
results of a random scan with or without using the freedom in the
matrices $T$ and $K$. The correlation that would be found in the
simplified scan (in black) is not found in a more general exploration
(in purple). Thanks to the master parametrization one can run
completely general scans and avoid finding this sort of \emph{fake}
correlations.

\section{Summary}
\label{sec:summary}

The master parametrization \cite{Cordero-Carrion:2018xre} can be
applied to any Majorana neutrino mass model and allows one to explore
its parameter space in a complete way and in full agreement with
neutrino oscillation data. Here we have detailed its ingredients and
illustrated its use for the particular case of the BNT model. Given
the large number of Majorana mass models in the literature, the master
parametrization constitutes a useful and general tool that allows one
to run systematic and automatizable phenomenological analyses in a
wide variety of scenarios beyond the SM.

\subsection*{Acknowledgements}

Work supported by the Spanish grants AYA2015-66899-C2-1-P,
SEV-2014-0398 and FPA2017-85216-P (AEI/FEDER, UE), PROMETEO/2018/165
and SEJI/2018/033 (Generalitat Valenciana) and the Spanish Red
Consolider MultiDark FPA2017-90566-REDC.

\section*{References}

\providecommand{\newblock}{}


\begin{thebibliography}{10}
\expandafter\ifx\csname url\endcsname\relax
  \def\url#1{{\tt #1}}\fi
\expandafter\ifx\csname urlprefix\endcsname\relax\def\urlprefix{URL }\fi
\providecommand{\eprint}[2][]{\url{#2}}

\bibitem{Ma:2016mwh}
Ma E and Popov O 2017 {\em Phys. Lett.\/} {\bf B764} 142--144
  (\textit{Preprint} \eprint{1609.02538})

\bibitem{CentellesChulia:2018gwr}
Centelles~Chuli\'a S, Srivastava R and Valle J~W~F 2018 {\em Phys. Lett.\/}
  {\bf B781} 122--128 (\textit{Preprint} \eprint{1802.05722})

\bibitem{Ma:1998dn}
Ma E 1998 {\em Phys. Rev. Lett.\/} {\bf 81} 1171--1174 (\textit{Preprint}
  \eprint{hep-ph/9805219})

\bibitem{Cai:2017jrq}
Cai Y, Herrero-Garc\'ia J, Schmidt M~A, Vicente A and Volkas R~R 2017 {\em
  Front.in Phys.\/} {\bf 5} 63 (\textit{Preprint} \eprint{1706.08524})

\bibitem{Cepedello:2018rfh}
Cepedello R, Fonseca R~M and Hirsch M 2018 {\em JHEP\/} {\bf 10} 197
  (\textit{Preprint} \eprint{1807.00629})

\bibitem{Boucenna:2014zba}
Boucenna S~M, Morisi S and Valle J~W~F 2014 {\em Adv. High Energy Phys.\/} {\bf
  2014} 831598 (\textit{Preprint} \eprint{1404.3751})

\bibitem{Anamiati:2018cuq}
Anamiati G, Castillo-Felisola O, Fonseca R~M, Helo J~C and Hirsch M 2018 {\em
  JHEP\/} {\bf 12} 066 (\textit{Preprint} \eprint{1806.07264})

\bibitem{Babu:2009aq}
Babu K~S, Nandi S and Tavartkiladze Z 2009 {\em Phys. Rev.\/} {\bf D80} 071702
  (\textit{Preprint} \eprint{0905.2710})

\bibitem{Cordero-Carrion:2018xre}
Cordero-Carri\'on I, Hirsch M and Vicente A 2018  (\textit{Preprint}
  \eprint{1812.03896})

\bibitem{future}
Cordero-Carri\'on I, Hirsch M and Vicente A  (in preparation)

\bibitem{Mohapatra:1986bd}
Mohapatra R~N and Valle J~W~F 1986 {\em Phys. Rev.\/} {\bf D34} 1642

\bibitem{Ma:2006km}
Ma E 2006 {\em Phys. Rev.\/} {\bf D73} 077301 (\textit{Preprint}
  \eprint{hep-ph/0601225})

\bibitem{Casas:2001sr}
Casas J~A and Ibarra A 2001 {\em Nucl. Phys.\/} {\bf B618} 171--204
  (\textit{Preprint} \eprint{hep-ph/0103065})

\bibitem{deSalas:2017kay}
de~Salas P~F, Forero D~V, Ternes C~A, Tortola M and Valle J~W~F 2018 {\em Phys.
  Lett.\/} {\bf B782} 633--640 (\textit{Preprint} \eprint{1708.01186})

\bibitem{Staub:2013tta}
Staub F 2014 {\em Comput. Phys. Commun.\/} {\bf 185} 1773--1790
  (\textit{Preprint} \eprint{1309.7223})

\bibitem{Porod:2011nf}
Porod W and Staub F 2012 {\em Comput. Phys. Commun.\/} {\bf 183} 2458--2469
  (\textit{Preprint} \eprint{1104.1573})

\bibitem{Porod:2014xia}
Porod W, Staub F and Vicente A 2014 {\em Eur. Phys. J.\/} {\bf C74} 2992
  (\textit{Preprint} \eprint{1405.1434})

\end{thebibliography}
\end{document}